\documentclass[amscd,amssymb,verbatim,12pt]{amsart}

\usepackage{epsfig}
\usepackage{url}
\usepackage[graph]{xy}

\newtheorem{theorem}{Theorem}[section]

\newtheorem*{note}{Note}

\theoremstyle{definition}

\newtheorem{example}[theorem]{Example}

\newtheorem{rules}{Rule}[section]

\theoremstyle{remark}
\newtheorem{remark}[theorem]{Remark}

\newcommand{\pFq}[5]{\ensuremath{{}_{#1}F_{#2} \left( \genfrac{}{}{0pt}{}{#3}{#4} \bigg| {#5} \right)}}
\newcommand{\bracket}[1]{\ensuremath{\left\langle {#1} \right\rangle}}
\newcommand{\mathd}{d}
\newcommand{\intd}{\,d}
\newcommand{\ift}{\int_{0}^{\infty}}
\newcommand{\eps}{\epsilon}

\begin{document}

\title[Method of brackets]{The method of brackets. Part 2: examples and
applications}

\author{Ivan Gonzalez}
\address{Departmento de Fisica y Centro de Estudios Subatomicos, 
Universidad Santa Maria, Valparaiso, Chile}
\email{ivan.gonzalez@usm.cl}

\author{Victor H. Moll}
\address{Department of Mathematics,
Tulane University, New Orleans, LA 70118}
\email{vhm@math.tulane.edu}

\author{Armin Straub}
\address{Department of Mathematics,
Tulane University, New Orleans, LA 70118}
\email{astraub@math.tulane.edu}



\numberwithin{equation}{section}

\begin{abstract}
A new heuristic method for the evaluation of definite integrals is presented.
This {\em method of brackets} has its origin in methods developed for the
evaluation of Feynman diagrams. The operational rules are described and
the method is illustrated with several examples.  The method of brackets reduces
the evaluation of a large class of definite integrals to the solution of a
linear system of equations. 
\end{abstract}

\maketitle

\section{Introduction}\label{sec-intro}

The {\em method of brackets} presented here provides a method for the
evaluation of a large class of definite integrals. The ideas were originally
presented in \cite{gonzalez-2007} in the context of integrals arising from
Feynman diagrams. A complete description of the operational rules of the method
together with a variety of examples was first discussed in \cite{gonzalez-moll1}.

The method is quite simple to work with and many of the entries from the
classical table of integrals \cite{gr} can be derived using this method.  The
basic idea is to introduce the formal symbol $\bracket{a}$, called a
\emph{bracket}, which represents the divergent integral
\begin{equation}\label{eq:bracketint}
  \ift x^{a -1} \intd x.
\end{equation}
The formal rules for operating with these brackets are described in Section
\ref{sec-bracket} and their justification (especially of the heuristic Rule
\ref{rule:freeparams}) is work-in-progress.  In particular, convergence issues
are ignored at the moment.  Roughly, each integral generates a linear system of
equations and for each choice of free variables the method yields a series with
the free variables as summation indices.  A heuristic rule states that those
converging in a common region give the desired evaluation. 

Section \ref{sec-gr} illustrates the method by evaluating the Laplace transform
of the Bessel function $J_{\nu}(x)$.  In this example, the two resulting series
converge in different regions and are analytic continuations of each other.
This is a general phenomenon which is used in Section \ref{sec-anal} to produce
an explicit analytic continuation of the hypergeometric function
$_{q+1}F_{q}(x)$.  Section \ref{sec-box} presents the evaluation of a family of
integrals $C_{n}$ appearing in Statistical Mechanics. These were introduced in
\cite{bailey2006a} as a toy model and their physical interpretation was
discovered later. The method of brackets is employed here to evaluate the first
four values, the only known cases (an expression for the next value $C_{5}$ in
terms of a double hypergeometric series is possible but is not given here).
The last section employs the method of brackets to resolve a Feynman diagram.

\section{The method of brackets}\label{sec-bracket}

The method of brackets discussed in this paper is based on the assignment of
the formal symbol $\bracket{a}$ to the divergent integral
(\ref{eq:bracketint}).
\begin{example}
If $f$ is given by the formal power series 
\begin{equation*}
  f(x) = \sum_{n=0}^{\infty} a_{n}x^{\alpha n+\beta-1},
\end{equation*}
then the improper integral of $f$ over the positive real axis is formally
written as the {\em bracket series}
\begin{equation}\label{eq:bracketserieseg}
  \ift f(x) \intd x = \sum_{n} a_{n} \bracket{\alpha n + \beta}.
\end{equation}
Here, and in the sequel, $\sum_n$ is used as a shorthand for 
$\sum_{n=0}^\infty$.
\end{example}

Formal rules for operating with brackets are described next. In particular,
Rule \ref{rule:evalbrackets} describes how to evaluate a bracket series such as
the one appearing in 
(\ref{eq:bracketserieseg}). To this end, it is useful to introduce the symbol
\begin{equation}\label{phi-def}
  \phi_{n} = \frac{(-1)^{n}}{\Gamma(n+1)},
\end{equation}
which is called the {\em indicator of} $n$.

\begin{example}
The gamma function has the bracket expansion
\begin{equation}
  \Gamma(a) = \ift x^{a-1} e^{-x} \intd x = \sum_{n} \phi_{n} \bracket{n + a}.
\end{equation}
\end{example}

\begin{rules}\label{rule:binom}
The bracket expansion
\begin{equation}
  \frac{1}{(a_{1} + a_{2} + \cdots + a_{r})^{\alpha}}
  = \sum_{m_1,\ldots,m_r} \phi_{m_1,\ldots,m_r} a_1^{m_1} \cdots a_r^{m_r}
  \frac{\bracket{\alpha + m_1+\cdots+m_r}}{\Gamma(\alpha)}
\end{equation}
holds.  Here $\phi_{m_1,\ldots,m_r}$ is a shorthand notation for the product
$\phi_{m_{1}} \cdots \phi_{m_{r}}$. If there is no possibility of confusion
this will be further abridged as $\phi_{\{m\}}$. The notation $\sum_{\{m\}}$ is
to be understood likewise.
\end{rules}

\begin{rules}\label{rule:evalbrackets}
A series of brackets is assigned a value according to
\begin{equation}
  \sum_{n} \phi_{n} f(n) \bracket{ a n + b} = 
  \frac{1}{|a|} f(n^{\ast}) \Gamma(-n^{\ast}),
\end{equation}
where $n^\ast$ is the solution of the equation $an+b = 0$. Observe that this
might result in the replacing of the index $n$, initially a nonnegative
integer, by a complex number $n^\ast$.

Similarly, a higher dimensional bracket series, that is,
\begin{equation}
  \sum_{\{n\}} \phi_{\{n\}} f(n_{1},\ldots,n_{r}) 
  \bracket{ a_{11}n_{1}+ \cdots a_{1r}n_{r} + c_{1} } \cdots 
  \bracket{ a_{r1}n_{1}+ \cdots a_{rr}n_{r} + c_{r} } \nonumber
\end{equation}
is assigned the value 
\begin{equation}
  \frac{1}{| \text{det}(A) |} f(n_{1}^{*}, \cdots, n_{r}^{*}) 
  \Gamma(-n_{1}^\ast) \cdots \Gamma(-n_{r}^\ast),
\end{equation}
where $A$ is the matrix of coefficients $(a_{ij})$ and $(n_i^\ast)$ is the
solution of the linear system obtained by the vanishing of the brackets.  The
value is not defined if the matrix $A$ is not invertible.
\end{rules}

\begin{rules}\label{rule:freeparams}
In the case where a higher dimensional series has more summation
indices than brackets, the appropriate number of free variables is chosen among
the indices.  For each such choice, Rule \ref{rule:evalbrackets}
yields a series.  Those converging  in a common region are added to
evaluate the desired integral. 
\end{rules}

\section{An example from Gradshteyn and Ryzhik}\label{sec-gr}

The second author is involved in a long term project of providing 
proofs of all the entries from the classical table of integrals by 
Gradshteyn and Ryzhik \cite{gr}. The proofs can be found at:
\begin{center}
\url{http://www.math.tulane.edu/~vhm/Table.html}
\end{center}

In this section the method of brackets is illustrated to find
\begin{equation}\label{eq:besselint}
  \ift x^{\nu} e^{- \alpha x} J_{\nu}(\beta x ) \intd x =
  \frac{(2 \beta)^{\nu} \Gamma( \nu + \tfrac{1}{2} )}{\sqrt{\pi} 
  ( \alpha^{2} + \beta^{2})^{\nu + 1/2}}
\end{equation}
which is entry $6.623.1$ of \cite{gr}. Here
\begin{equation}
J_{\nu}(x) = \sum_{k=0}^{\infty} \frac{(-1)^{k} (x/2)^{2k+\nu}}{k! \, 
\Gamma(k + \nu + 1)}
\end{equation}
is the Bessel function of order $\nu$. To this end, the integrand is expanded as
\begin{align}
  e^{- \alpha x} J_{\nu}(\beta x ) & = \left( \sum_{n} \phi_n (\alpha x)^n \right)
  \left( \sum_{k} \phi_k \frac{(\tfrac{\beta x}{2})^{2k+\nu}}{\Gamma(k+\nu+1)} \right) \\
  & = \sum_{k,n} \phi_{k,n} \frac{\alpha^n (\tfrac{\beta}{2})^{2k+\nu}}{\Gamma(k+\nu+1)} x^{n+2k+2\nu}, \nonumber
\end{align}
so as to obtain the bracket series
\begin{equation}\label{eq:besselbracketseries}
  \ift e^{- \alpha x} J_{\nu}(\beta x ) \mathd x = 
  \sum_{k,n} \phi_{k,n} \frac{\alpha^n (\tfrac{\beta}{2})^{2k+\nu}}{\Gamma(k+\nu+1)} \bracket{n+2k+2\nu+1}.
\end{equation}
The evaluation of this double sum by the method of brackets produces two series
corresponding to using either $k$ or $n$ as the free variable when applying Rule \ref{rule:evalbrackets}.

\subsection*{The index $k$ is free}
Choosing $k$ as the free variable when applying Rule \ref{rule:evalbrackets} to
(\ref{eq:besselbracketseries}), yields $n^\ast = -2k-2\nu-1$ and thus the
resulting series
\begin{align}\label{eq:besselintserieskfree}
  & \sum_{k} \phi_{k} \frac{\alpha^{-2k-2\nu-1} (\tfrac{\beta}{2})^{2k+\nu}}{\Gamma(k+\nu+1)} \Gamma(2k+2\nu+1) \\
  = {} & \alpha^{-2\nu-1} (\tfrac{\beta}{2})^\nu \frac{\Gamma(2\nu+1)}{\Gamma(\nu+1)} \pFq10{\nu+\tfrac12}{-}{-\frac{\beta^2}{\alpha^2}}. \nonumber
\end{align}
The right-hand side employs the usual notation for the hypergeometric function
\begin{equation}
  \pFq{p}{q}{a_1,\ldots,a_p}{b_1,\ldots,b_q}{x}
  = \sum\limits_{n=0}^{\infty }\frac{\left( a_{1}\right)
  _{n} \cdots \left( a_{p}\right) _{n}}{\left( b_{1}\right) _{n} \cdots \left(
  b_{q}\right) _{n}}\frac{x^{n}}{n!}
\end{equation}
where $(\alpha)_{n} = \tfrac{\Gamma(\alpha + n)}{\Gamma(\alpha)}$ is the
Pochhammer symbol.  Note that the ${}_1F_0$ in (\ref{eq:besselintserieskfree})
converges provided $|\beta| < |\alpha|$.  In this case, the standard
identity ${}_1F_0(a|x) = (1-x)^{-a}$ together with the duplication formula for
the $\Gamma$ function shows that the series in (\ref{eq:besselintserieskfree})
is indeed equal to the right-hand side of (\ref{eq:besselint}).

\subsection*{The index $n$ is free}
In this second case, the linear system in Rule \ref{rule:evalbrackets} has
determinant $2$ and yields $k^\ast = -n/2-\nu-1/2$.  This gives 
\begin{equation}\label{eq:besselintseriesnfree}
  \frac{1}{2} \sum_{n} \phi_{n} \frac{\alpha^{n} (\tfrac{\beta}{2})^{-n-\nu-1}}{\Gamma(-n/2+1/2)} \Gamma(n/2+\nu+1/2).
\end{equation}
This series now converges provided that $|\beta | > |\alpha|$ in which case it
again sums to the right-hand side of (\ref{eq:besselint}).

\begin{note}
This is the typical behavior of the method of brackets. The
different choices of indices as free variables give representations of the
solution valid in different regions. Each of these is an analytic continuation
of the other ones.
\end{note}

\section{Integrals of the {I}sing class}\label{sec-box}

In this section the method of brackets is used to discuss the integral
\begin{equation}
  C_n = \frac{4}{n!} \ift \cdots \ift
  \frac{1}{\left( \sum_{j=1}^{n} ( u_{j} + 1/u_{j}) \right)^{2}}
  \, \frac{\mathd u_{1}}{u_{1}} \cdots \frac{\mathd u_{n}}{u_{n}}.
\end{equation}
This family was introduced in \cite{bailey2006a} as a caricature of the 
{\emph{Ising susceptibility integrals}}
\begin{equation}\label{eq:Dn}
  D_{n} = \frac{4}{n!} 
  \ift \cdots \ift 
  \prod_{i<j} \left( \frac{u_{i}-u_{j}}{u_{i}+u_{j}} \right)^{2} \, 
  \frac{1}{\left( \sum_{j=1}^{n} ( u_{j} + 1/u_{j}) \right)^{2} }
  \, \frac{\mathd u_{1}}{u_{1}} \cdots \frac{\mathd u_{n}}{u_{n}}.
\end{equation}
Actually, the integrals $C_{n}$ appear naturally in the analysis of certain
amplitude transforms \cite{palmer-tracy}. The first few values are given by
\begin{equation}\label{eq:Cnvalues}
  C_{1} = 2, \; C_{2} = 1, \; C_{3} = L_{-3}(2), \; C_{4} = \frac{7}{12} \zeta(3).
\end{equation}
Here, $L_D$ is the Dirichlet L-function. In this case,
\begin{equation}
  L_{-3}(2) = \sum_{n=0}^{\infty} \left( \frac{1}{(3n+1)^2} - \frac{1}{(3n+2)^2} \right).
\end{equation}
No analytic expression for $C_n$ is known for $n\ge5$. Similarly,
\begin{equation}\label{eq:Dnvalues}
  D_{1} = 2, \; D_{2} = \frac{1}{3}, \; D_{3} = 8 + 
  \frac{4 \pi^{2}}{3} - 27L_{-3}(2), \;
  D_{4} = \frac{4 \pi^{2}}{9} - \frac{1}{6} - \frac{7}{12} \zeta(3)
\end{equation}
are given in \cite{bailey2006a}. High precision numerical evaluation and PSLQ
experiments have further produced the conjecture
\begin{align}
  D_{5} = {} & 42 - 1984 \text{Li}_{4}(\tfrac{1}{2}) 
  + \frac{189}{10} \pi^{4} - 74 \zeta(3) 
  -1272 \zeta(3) \ln 2 + 40 \pi^{2} \ln^{2} 2 \\
  & - \frac{62}{3} \pi^{3} +
  \frac{40}{3} \pi^{2} \ln 2 + 88 \ln^{4}2 + 464 \ln^{2}2 - 40 \ln 2.
  \nonumber
\end{align}

The integral $C_{n}$ is the special case $k=1$ of the family 
\begin{equation}
  C_{n,k} = \frac{4}{n!} \ift \cdots \ift
  \frac{1}{\left( \sum_{j=1}^{n} ( u_{j} + 1/u_{j}) \right)^{k+1}}
  \, \frac{\mathd u_{1}}{u_{1}} \cdots \frac{\mathd u_{n}}{u_{n}}
\end{equation}
that also gives the moments of powers of the Bessel function $K_0$ via
\begin{equation}
  C_{n,k} = \frac{2^{n-k+1}}{n! \, k!} c_{n,k} := \frac{2^{n-k+1}}{n! \, k!} \ift t^{k}K_{0}^{n}(t) \intd t.
\end{equation}
The values 
\begin{equation}\label{eq:c2k}
  c_{1,k} = 2^{k-1} \Gamma^{2} \left( \frac{k+1}{2} \right), \quad
  c_{2,k} = \frac{\sqrt{\pi}}{4} 
  \frac{\Gamma^{3} \left( \frac{k+1}{2} \right)}{\Gamma \left( \frac{k}{2} +1 \right)}, 
\end{equation}
as well as the recursion
\begin{equation}
  (k+1)^{4}c_{3,k} -2(5k^{2}+20k+21)c_{3,k+2} + 9c_{3,k+4}=0
\end{equation}
with initial data 
\begin{equation}\label{init-case3}
  c_{3,0} = \frac{3 \alpha}{32 \pi}, \;
  c_{3,1} = \frac{3}{4} L_{-3}(2), \;
  c_{3,2} = \frac{\alpha}{96 \pi} - \frac{4 \pi^{5}}{9 \alpha}, \;
  c_{3,3} = L_{-3}(2) - \frac{2}{3},
\end{equation}
where $\alpha = 2^{-2/3} \Gamma^{6}(\tfrac{1}{3})$ are given in
\cite{borwein-2008b} and \cite{borwein-2008a}. 

The evaluation of these integrals presented in the literature usually 
begins with the introduction of 
spherical coordinates. This reduces the dimension of $C_{n}$ by two and 
immediately gives the values of $C_{1}$ and $C_{2}$. The evaluation of
$C_{3}$ is reduced to the logarithmic integral
\begin{equation}
  C_{3} = \frac{2}{3} \int_{0}^{\infty} \frac{\ln(1+x) \intd x }{x^{2}+x+1}.
\end{equation}
Its value is obtained by the change of variables $x \to \tfrac{1}{t}-1$ 
followed by an expansion of the integrand. 
A systematic discussion of these type of logarithmic integrals is provided 
in \cite{luis2}. The value of $C_{4}$ is obtained via the double integral 
representation
\begin{equation}
  C_{4} = \frac{1}{6} \ift \ift
  \frac{\ln(1+x+y) }{(1+x+y)(1+1/x+1/y)-1} 
  \frac{\mathd x}{x} \frac{\mathd y}{y}.
\end{equation}
Moreover, the limiting behavior
\begin{equation}
 \lim_{n \to \infty} C_{n} = 2e^{-2 \gamma}
\end{equation}
was established in \cite{bailey2006a}. 

In this section the method of brackets is used to obtain the expressions for
$C_{2}$, $C_{3}$, and $C_{4}$ described above.  An advantage of this method is
that it systematically gives an analytic expression for these integrals. When
applied to $C_5$,  the method produces a double series representation which
is not discussed here.

\subsection{Evaluation of $C_{2,k}$}

The numbers $C_{2,k}$ are given by
\begin{equation}\label{eq:C2k}
  C_{2,k} = 2 \ift \ift \frac{\mathd x \, \mathd y }{xy \, 
  \left( x+ 1/x+y+1/y\right)^{k+1}}.
\end{equation}
A direct application of the method of brackets, by applying Rule
\ref{rule:binom} to the integrand as in (\ref{eq:C2k}), results in a bracket
expansion involving a $4$-fold sum and $3$ brackets.  Rules
\ref{rule:evalbrackets} and \ref{rule:freeparams} translates this into a
collection of series with $4-3=1$ summation indices.  However, it is generally
desirable to minimize the final number of summations by reducing the number of
sums and increasing the number of brackets. In this example this is achieved by
writing
\begin{align*}
  C_{2,k} &= 2 \ift \ift
  \frac{(xy)^k \intd x \intd y}{( x^{2}y+y+xy^{2}+x) ^{k+1}} \\
  &= 2 \ift \ift \frac{(xy)^k \intd x \intd y}
  {(xy \left[ x+y\right] +\left[ x+y\right] ) ^{k+1}}.
\end{align*}
In the evaluation of these expressions, the term $(x+y)$
must be expanded at the last step. The method of brackets now yields
\begin{equation*}
  \frac{1}{\left( xy\left[ x+y\right] +\left[ x+y\right] \right)
  ^{k+1}}= \sum_{n_1,n_2} \phi_{n_1,n_2}
  \;x^{n_{1}}y^{n_{1}}\left( x+y\right)
  ^{n_{1}+n_{2}}\frac{\bracket{ k+1+n_{1}+n_{2}}}{\Gamma(k+1)},
\end{equation*}
and the expansion of the term $(x+y)$ gives
\begin{equation*}
  \frac{1}{\left( x+y\right) ^{-n_{1}-n_{2}}}=
  \sum_{n_3,n_4} \phi_{n_3,n_4}
  \;x^{n_{3}}y^{n_{4}}\frac{\bracket{
  -n_{1}-n_{2}+n_{3}+n_{4}} }{\Gamma \left(
  -n_{1}-n_{2}\right) }.
\end{equation*}
Replacing in the integral produces the bracket expansion
\begin{align*}
  C_{2,k} = {} & 2 \sum_{\{n\}} \phi_{\{n\}}
  \frac{\bracket{k+1+n_{1}+n_{2}}}{\Gamma(k+1)}
  \frac{\bracket{-n_{1}-n_{2}+n_{3}+n_{4}}}{\Gamma(-n_1-n_2)} \\
  &\times \bracket{k+1+n_{1}+n_{3}} \bracket{k+1+n_{1}+n_{4}}.
\end{align*}
The value of this formal sum  is now obtained by
solving the linear system $k+1+n_{1}+n_{2} =0$, $-n_{1}-n_{2}+n_{3}+n_{4}=0$,
$k+1+n_{1}+n_{3}=0$, and $k+1+n_{1}+n_{4} = 0$ coming from the vanishing of
brackets.  This system has determinant $2$ and its unique solution is
$n_{1}^{\ast }=n_{2}^{\ast }=n_{3}^{\ast }=n_{4}^{\ast }=-\tfrac{k+1}{2}$. It
follows that
\begin{equation}
  C_{2,k} = \frac{\Gamma \left( -n_{1}^{\ast }\right) \Gamma \left(
  -n_{2}^{\ast }\right) \Gamma \left( -n_{3}^{\ast }\right) \Gamma \left(
  -n_{4}^{\ast }\right) }{\Gamma(k+1) \Gamma \left( -n_{1}^{\ast }-n_{2}^{\ast }\right) }
  = \frac{\Gamma\left(\tfrac{k+1}{2}\right)^4}{\Gamma(k+1)^2}.
\end{equation}
Note that, upon employing Legendre's duplication formula for the $\Gamma$
function, this evaluation is equivalent to (\ref{eq:c2k}).  In particular, this
confirms the value $C_{2} = C_{2,1} = 1$ in (\ref{eq:Cnvalues}).

\begin{remark}
The evaluation
\begin{align}
  C_{2,k}(\alpha,\beta) &= 2 \ift\ift \frac{x^{\alpha-1} y^{\beta-1} \intd x \intd y }{\left( x+ 1/x+y+1/y\right) ^{k+1}}\\
  & = \frac{\Gamma\left(\tfrac{k+1+\alpha+\beta}{2}\right)\Gamma\left(\tfrac{k+1-\alpha-\beta}{2}\right)
  \Gamma\left(\tfrac{k+1+\alpha-\beta}{2}\right)\Gamma\left(\tfrac{k+1-\alpha+\beta}{2}\right)}{\Gamma(k+1)^2} \nonumber
\end{align}
that generalizes $C_{2,k}$ is obtained as a bonus. Similarly, 
\begin{align}
  J_{r,s}(\alpha,\beta) & = 2 \ift\ift \frac{x^{\alpha-1} y^{\beta-1} \intd x \intd y }{(x+y)^r (xy+1)^s}\\
  & = \frac{\Gamma\left(\tfrac{-r+\alpha+\beta}{2}\right)\Gamma\left(\tfrac{2s+r-\alpha-\beta}{2}\right)
  \Gamma\left(\tfrac{r+\alpha-\beta}{2}\right)\Gamma\left(\tfrac{r-\alpha+\beta}{2}\right)}{\Gamma(r)\Gamma(s)}. \nonumber
\end{align}
Note that $C_{2,k}(\alpha,\beta) = J_{k+1,k+1}(\alpha + k+1, \beta + k+1)$.
\end{remark}

\begin{remark}
The Ising susceptibility integral $D_2$, see (\ref{eq:Dn}), is obtained
directly from the expression for $J_{r,s}$ given above. Indeed,
\begin{align}
  D_2 & = 2 \ift\ift (x^2-2xy+y^2) \frac{x y \intd x \intd y}{(x+y)^4 (xy+1)^2}\\
  & = 2 \left( J_{4,2}(4,2) -2J_{4,2}(3,3) + J_{4,2}(2,4) \right) \nonumber\\
  & = \frac{1}{3}. \nonumber
\end{align}
This agrees with (\ref{eq:Dnvalues}). This technique also yields the 
generalization
\begin{align}
  D_2(\alpha,\beta) & = 2 \ift\ift \left(\frac{x-y}{x+y}\right)^2 \frac{x^{\alpha-1} y^{\beta-1} \intd x \intd y}{\left(x+1/x+y+1/y\right)^2}\\
  & = \frac{(b-a)(b+a)(2+(b-a)^2) \pi^2}{12(\cos(\alpha\pi) - \cos(\beta\pi))} \nonumber
\end{align}
with limiting case 
$D_2(\alpha,\alpha) = \tfrac13 \tfrac{\alpha\pi}{\sin(\alpha\pi)}$.
\end{remark}

\subsection{Evaluation of $C_{3,k}$}

Next, consider the integral
\begin{align}\label{eq:C3k}
  C_{3,k} & = \frac{2}{3} \ift\ift\ift \frac{\intd x \intd y \intd z}{xyz \, 
  \left( x+ 1/x+y+1/y + z + 1/z \right)^{k+1}} \\
  & = \frac{2}{3} \ift\ift\ift \frac{(xyz)^k \intd x \intd y \intd z}{\left( xyz\left( x+y\right) +
  z\left( x+y\right) +xyz^{2}+xy\right) ^{k+1}}. \nonumber
\end{align}
The second form of the integrand is motivated by the desire to 
to minimize the number of sums and to maximize the number of
brackets in the expansion. The denominator is now expanded as
\begin{equation*}
  \sum_{\{n\}} \phi_{\{n\}}
  (xy)^{n_{1}+n_{3}+n_{4}} z^{n_{1}+n_{2}+2n_{3}}
  \left( x+y\right) ^{n_{1}+n_{2}}\frac{\bracket{k+1
  +n_{1}+n_{2}+n_{3}+n_{4}}}{\Gamma \left( k+1 \right)},
\end{equation*}
and further expanding $(x+y)^{n_1+n_2}$ as
\begin{equation*}
  \left( x+y\right) ^{n_{1}+n_{2}}= \sum_{n_5,n_6} \phi_{n_5,n_6}
  \;x^{n_{5}}y^{n_{6}}\frac{\bracket{
  -n_{1}-n_{2}+n_{5}+n_{6}} }{\Gamma(-n_{1}-n_{2})}
\end{equation*}
produces a complete bracket expansion of the integrand of $C_{3,k}$.
Integration then yields
\begin{align}\label{eq:C3kbrackets}
  C_{3,k} = {} & \frac{2}{3} \frac{1}{k!} \sum_{\{n\}} \phi_{\{n\}}\;
    \frac{\bracket{-n_{1}-n_{2}+n_{5}+n_{6}} }{\Gamma \left( -n_{1}-n_{2}\right) }\\
  & \times \bracket{k+1 +n_{1}+n_{2}+n_{3}+n_{4}}
    \bracket{k+1+n_{1}+n_{3}+n_{4}+n_{5}} \nonumber\\
  & \times \bracket{k+1+n_{1}+n_{3}+n_{4}+n_{6}} \bracket{k+1+n_{1}+n_{2}+2n_{3}}.\nonumber
\end{align}
This expression is regularized by replacing the bracket $\langle
k+1+n_{1}+n_{2}+2n_{3}\rangle$ with $\langle
k+1+n_{1}+n_{2}+2n_{3}+\epsilon\rangle$ with the intent of letting $\epsilon
\rightarrow 0$. (This corresponds to multiplying the initial integrand with
$z^\epsilon$; however, note that many other regularizations are possible and
eventually lead to Theorem \ref{thm:I3k}. It will become clear shortly, see
(\ref{eq:I3kterm}), why regularizing is necessary.) The method of brackets now
gives a set of series expansions obtained by the vanishing of the five brackets
in (\ref{eq:C3kbrackets}). The solution of the corresponding linear system
(which has determinant $2$) leaves one free index and produces the integral as
a series in this variable. Of the six possible free indices, only $n_{3}$ and
$n_{4}$ produce convergent series (more specifically, for each free index one
obtains a hypergeometric series ${}_3F_2$ times an expression free of the
index; for the indices $n_3,n_4$ the argument of this ${}_3F_2$ is
$\tfrac{1}{4}$ while otherwise it is $4$.) The heuristic Rule
\ref{rule:freeparams} states that their sum yields the value of the integral:
\begin{equation}
  C_{3,k} = \frac13 \lim_{\epsilon \rightarrow 0} \frac{1}{k!}\sum_{n=0}^{\infty} \frac{(-1)^n}{n!} (f_{k,n}(\epsilon) + f_{k,n}(-\epsilon))
\end{equation}
where
\begin{equation}\label{eq:I3kterm}
  f_{k,n}(\epsilon) = \frac{\Gamma\left(n+\tfrac{k+1+\epsilon}{2}\right)^4 \Gamma(-n-\epsilon)}{\Gamma(2n+k+1+\epsilon)}.
\end{equation}
Observe that the terms $f_{k,n}(\epsilon)$ are contributed by the index $n_3$
while the terms $f_{k,n}(-\epsilon)$ come from the index $n_4$. At $\epsilon =
0$, each of them has a simple pole. Consequently, the even
combination $f_{k,n}(\epsilon) + f_{k,n}(-\epsilon)$ has no pole at $\epsilon =
0$.  Using the expansions
\begin{equation}
  \Gamma(x+\epsilon) = \Gamma(x)(1 + \psi(x)\epsilon) + O(\epsilon^2),
\end{equation}
for $x \neq 0,-1,-2,\ldots$, as well as
\begin{equation}
  \Gamma(-n+\epsilon) = \frac{(-1)^n}{n!} \left(\frac{1}{\epsilon} + \psi(n+1)\right) + O(\epsilon),
\end{equation}
for $n = 0,1,2,\ldots$, provides the next result.

\begin{theorem}\label{thm:I3k}
The integrals $C_{3,k}$ are given by
\begin{equation*}
  C_{3,k} = \frac{2}{3} \frac{1}{k!} \sum_{n=0}^{\infty} \frac{1}{(n!)^2} \frac{\Gamma\left(n+\tfrac{k+1}{2}\right)^4}{\Gamma(2n+k+1)}
    \left( \psi(n+1) - 2\psi\left(n+\tfrac{k+1}{2}\right) + \psi(2n+k+1) \right).
\end{equation*}
\end{theorem}

In particular, for $k=1$
\begin{equation}
  C_3 = \frac{2}{3} \sum_{n=0}^{\infty} \frac{(n!)^2}{(2n+1)!}
    \left(\psi(2n+2) - \psi(n+1) \right).
\end{equation}
The evaluation of this sum using \texttt{Mathematica 7} yields a large
collection of special values of (poly-)logarithms. After simplifications, it
yields $C_3 = L_{-3}(2)$ as in (\ref{eq:Cnvalues}).

\begin{remark}
An extension of Theorem \ref{thm:I3k} is presented next:
\begin{equation}
  C_{3,k}(\alpha,\beta,\gamma) = \ift\ift\ift 
  \frac{x^{\alpha-1} y^{\beta-1} z^{\gamma-1} \intd x \intd y \intd z}{\left(x+ 1/x+y+1/y + z + 1/z \right)^{k+1}},
\end{equation}
for $\gamma=0$, is given by
\begin{equation*}
  \frac{1}{k!} \sum_{n=0}^{\infty} \frac{1}{(n!)^2} \frac{\Gamma\left(n+\tfrac{k+1 \pm \alpha \pm \beta}{2}\right)}{\Gamma(2n+k+1)}
    \left( \psi(n+1) - \frac12 \psi\left(n+\tfrac{k+1 \pm \alpha \pm \beta}{2}\right) + \psi(2n+k+1) \right)
\end{equation*}
where the notation $\Gamma(n+\tfrac{k+1 \pm \alpha \pm \beta}{2}) =
\Gamma(n+\tfrac{k+1+\alpha+\beta}{2})\Gamma(n+\tfrac{k+1+\alpha-\beta}{2})\cdots$
as well as $\psi(n+\tfrac{k+1 \pm \alpha \pm \beta}{2}) =
\psi(n+\tfrac{k+1+\alpha+\beta}{2}) + \psi(n+\tfrac{k+1+\alpha-\beta}{2}) +
\cdots$ is employed.  Similar expressions can be given for other integral
values of $\gamma$. In the case where $\gamma$ is not integral,
$C_{3,k}(\alpha,\beta,\gamma)$ can be written as a sum of two ${}_3F_2$'s with
$\Gamma$ factors.  The symmetry of $C_{3,k}(\alpha,\beta,\gamma)$ in
$\alpha,\beta,\gamma$, shows that this can be done if at least one of these
arguments is nonintegral.
\end{remark}

\subsection{Evaluation of $C_{4}$}

The last example discussed here is
\begin{equation*}
  C_{4} = \frac16 \ift\ift\ift\ift \frac{\intd x \intd y \intd z \intd w}{xyzw
  \left( x+1/x+y+1/y+z+1/z+w+1/w\right) ^{2}}.
\end{equation*}
To minimize the number of sums and to maximize the number of brackets this is
rewritten as
\begin{equation*}
  \frac16 \ift\ift\ift\ift \frac{x^{1+\eps}y^{1+\eps}z^{1+\eps}w^{1+ \eps} \intd x \intd y \intd z \intd w}
  {\left[ Axyzw (x+y) +zw(x+y) +xyzw( z+w) +xy(z+w) \right]^{2}}
\end{equation*}
with the intent of letting $\eps \to 0$ and $A \to 1$.  As in the case of
$C_{3,k}$, the {\em regulator} parameter $\eps$ is introduced to cure the
divergence of the resulting expressions.  Similarly, the parameter $A$ is
employed to divide the resulting sums into convergence groups according to
the heuristic Rule \ref{rule:freeparams}.  The denominator expands as
\begin{align*}
  \sum_{\{n\}} & \phi_{\{n\}}
  \;A^{n_{1}}x^{n_{1}+n_{3}+n_{4}}y^{n_{1}+n_{3}+n_{4}}z^{n_{1}+n_{2}+n_{3}}
  w^{n_{1}+n_{2}+n_{3}} \\
  & \times (x+y)^{n_{1}+n_{2}} (z+w)^{n_{3}+n_{4}}
  \bracket{2+n_{1}+n_{2}+n_{3}+n_{4}}.
\end{align*}
As before,
\begin{equation*}
  (x+y)^{n_{1}+n_{2}} = \sum_{n_5,n_6} \phi_{n_5,n_6}\;x^{n_{5}}y^{n_{6}}\frac{\bracket{
  -n_{1}-n_{2}+n_{5}+n_{6}}}{\Gamma(-n_{1}-n_{2})}
\end{equation*}
and 
\begin{equation*}
  (z+w)^{n_{3}+n_{4}} = \sum_{n_7,n_8} \phi_{n_7,n_8}\;z^{n_{7}}w^{n_{8}}\frac{\bracket{
  -n_{3}-n_{4}+n_{7}+n_{8}}}{\Gamma(-n_{3}-n_{4})}.
\end{equation*}
These expansions of the integrand yield the bracket series
\begin{align}
  \frac16 \sum_{\{n\}} & \phi_{\{n\}} \; A^{n_{1}} \bracket{2+n_{1}+n_{2}+n_{3}+n_{4}} \\
  & \times \; \frac{\bracket{-n_{1}-n_{2}+n_{5}+n_{6}}}{\Gamma(-n_{1}-n_{2})}
  \frac{\bracket{-n_{3}-n_{4}+n_{7}+n_{8}}}{\Gamma(-n_{3}-n_{4})} \nonumber \\
  & \times \; \bracket{2+\epsilon+n_{1}+n_{3}+n_{4}+n_{5}}
  \bracket{2+\epsilon+n_{1}+n_{3}+n_{4}+n_{6}} \nonumber \\
  & \times \; \bracket{2+\epsilon+n_{1}+n_{2}+n_{3}+n_{7}}
  \bracket{2+\epsilon +n_{1}+n_{2}+n_{3}+n_{8}}. \nonumber
\end{align}
The evaluation of this bracket series by Rules \ref{rule:evalbrackets} and
\ref{rule:freeparams} yields hypergeometric series with arguments $A$ ($n_1$,
$n_2$, $n_5$, or $n_6$ chosen as the free index) and $1/A$ ($n_3$, $n_4$,
$n_7$, or $n_8$ chosen as the free index).  Either combination produces an
expression for the integral $C_4$. Taking those with argument $A$ (the indices
$n_5$ and $n_6$ yield the same series; however, it is only taken into account
once) gives
\begin{align}
  \frac{1}{12} A^{-\eps} \Gamma^2(\eps) \Gamma^2(1-\eps) \bigg( &
  \frac{A^{\eps}}{1+2\eps} \pFq21{\tfrac12+\eps,1}{\tfrac32+\eps}{A} \\
  + & \frac{A^{-\eps}}{1-2\eps} \pFq21{\tfrac12-\eps,1}{\tfrac32-\eps}{A}
  - 2 \pFq21{\tfrac12,1}{\tfrac32}{A} \bigg). \nonumber
\end{align}
As $\eps \to 0$, the limiting value is
\begin{align}
  \frac{1}{24}\ln^{2}A \, \ln \left( \frac{1+\sqrt{A}}{1-\sqrt{A}}\right)
  + & \frac{1}{3\sqrt{A}} \left[ \text{Li}_{3}(\sqrt{A}) - \text{Li}_{3}( -\sqrt{A}) \right] \\
  - & \frac{\ln A }{6\sqrt{A}}\left[ \text{Li}_{2}(\sqrt{A}) - \text{Li}_{2}(- \sqrt{A}) \right].
\nonumber
\end{align}
Finally, the value of $C_4$ is obtained by taking $A \to 1$: 
\begin{equation}
  C_{4} = \frac13 \left[ \text{Li}_{3}(1) - \text{Li}_{3}(-1) \right] = \frac{7}{12} \zeta(3).
\end{equation}
This agrees with (\ref{eq:Cnvalues}).

\section{Analytic continuation of hypergeometric functions}\label{sec-anal}

The hypergeometric function ${_{p}F_{q}}$, defined
by the series
\begin{equation}\label{pFqseries}
  {}_pF_q(x) = \pFq{p}{q}{a_1,\ldots,a_p}{b_1,\ldots,b_q}{x}
  = \sum\limits_{n=0}^{\infty} \frac{(a_{1})_{n} \cdots (a_{p})_{n}}
  {(b_{1})_{n} \cdots (b_{q})_{n}} \frac{x^{n}}{n!},
\end{equation}
converges for all $x \in \mathbb{C}$ if $p <  q+1$ and for $|x| < 1$ if
$p=q+1$. In the remaining case, $p > q+1$, the series diverges for $x \neq 0$.
The analytic continuation of the series $_{q+1}F_{q}$ has been recently
considered in \cite{skorokhodov1,skorokhodov2}. In this section a
brackets representation of the hypergeometric series is obtained and 
then employed to produce its analytic extension.

\begin{theorem}\label{thm:hypbrackets} 
The bracket representation of the hypergeometric function is given by
\begin{equation*}
  {}_pF_q(x) = \sum\limits_{\substack{n \\ t_{1},\ldots,t_{p} \\ s_{1},\ldots,s_{q}}}
  \phi _{n,\{t\},\{s\}}\ \left[(-1)^{q-1}x\right]^{n} \prod\limits_{j=1}^{p}
  \frac{\bracket{a_{j}+n+t_{j}}}{\Gamma(a_j)} \,
  \prod\limits_{k=1}^{q}\frac{\bracket{1-b_{k}-n+s_{k}}}{\Gamma(1-b_k)}.
\end{equation*}
\end{theorem}

\begin{proof}
This follows from (\ref{pFqseries}) and the representations
\begin{equation}
  (a_{j})_{n} = \frac{\Gamma(a_{j}+n)}{\Gamma(a_{j})}
  = \frac{1}{\Gamma(a_{j})} \ift \tau ^{a_{j}+n-1} e^{-\tau} \intd \tau
  = \sum_{t_{j}} \phi_{t_{j}} \frac{\bracket{a_{j}+n+t_{j}}}{\Gamma(a_{j})}
\end{equation}
as well as
\begin{equation}
  \frac{1}{(b_{k})_{n}}
  = (-1)^{n} \frac{\Gamma(1-b_{k}-n)}{\Gamma(1-b_{k})}
  = (-1)^{n} \sum_{s_{k}} \phi_{s_{k}} \, \frac{\bracket{1-b_{k}-n+s_{k}}}{\Gamma(1-b_{k})}
\end{equation}
for the Pochhammer symbol.
\end{proof}

The bracket expression for the hypergeometric function given in Theorem
\ref{thm:hypbrackets} contains $p+q$ brackets and $p+q+1$ indices ($n, \,
t_{j}$ and $s_{k}$). This leads to a full rank system
\begin{align}\label{sys-1}
  a_{j} + n + t_{j} & = 0 \qquad \text{ for } 1 \leq j \leq p \\
  1 - b_{k} - n + s_{k} & = 0 \qquad \text{ for } 1 \leq k \leq q. \nonumber
\end{align}
of linear equations of size $(p+q+1) \times (p+q)$ and determinant $1$. For
each choice of an index as a free variable the method of brackets yields a
one-dimensional series for the integral. 

\subsection*{Series with $n$ as a free variable}
Solving (\ref{sys-1}) yields $t_{j}^\ast = -a_{j} - n$ and $s_{k}^\ast =
-(1-b_{k}) + n$ with $1 \leq j \leq p$ and $1 \leq k \leq q$. Rule
\ref{rule:evalbrackets} yields
\begin{equation*}
  \sum_{n=0}^{\infty} 
  \frac{[ (-1)^{q}x]^{n}}{n!}
  \prod_{j=1}^{p} \frac{\Gamma(n+a_{j})}{\Gamma(a_j)} \prod_{k=1}^{q} \frac{\Gamma(-n+1-b_{k})}{\Gamma(1-b_k)}
  = \sum_{n=0}^{\infty} \frac{(a_{1})_{n} \cdots (a_{p})_{n}}
  {(b_{1})_{n} \cdots (b_{q})_{n}} \frac{x^{n}}{n!}.
\end{equation*}
This is the original series representation (\ref{pFqseries}) of the
hypergeometric function.  In particular, in the case $q = p-1$, this series
converges for $|x| < 1$.  

\subsection*{Series with $t_{i}$ as a free variable}
Fix an index $i$ in the range $1 \leq i \leq p$ and solve (\ref{sys-1}) to get
$n^\ast=-a_{i}-t_{i}$, as well as $t_{j}^\ast = t_{i}-a_{j}+a_{i}$ for $1 \leq
j \leq p$, $j \neq i$, and $s_{k}^\ast=-(1-b_{k}) -a_{i}-t_{i}$ for $1 \leq k
\leq q$.  The method of brackets then produces the series
\begin{equation*}
  \sum_{t_i} \phi_{t_i}\left[ (-1)^{q-1}x \right]^{-t_i-a_i} \frac{\Gamma(t_i+a_i)}{\Gamma(a_i)}
  \prod_{j \ne i} \frac{\Gamma(a_j-a_i-t_i)}{\Gamma(a_j)}
  \prod_{k} \frac{\Gamma(1-b_k+a_i+t_i)}{\Gamma(1-b_k)}
\end{equation*}
which may be rewritten as
\begin{align}\label{eq:hyptifree}
  (-x)^{-a_i} &
  \prod_{j \ne i} \frac{\Gamma(a_j-a_i)}{\Gamma(a_j)}
  \prod_{k} \frac{\Gamma(b_k)}{\Gamma(b_k-a_i)}\\
  & \times \; \pFq{q+1}{p-1}
  {a_i, \{1-b_k+a_i\}_{1\le k \le q}}
  {\{1-a_j+a_i\}_{1\le j \le p, j\ne i}}
  {\frac{(-1)^{p+q-1}}{x}}.
  \nonumber
\end{align}
Recall that the initial hypergeometric series ${}_pF_q(x)$ converges for some
$x\ne 0$ if and only if $p \le q+1$.  Hence, assuming that $p \le q+1$, observe
that the hypergeometric series (\ref{eq:hyptifree}) converges for some $x$ if
and only if $p = q+1$.

\subsection*{Series with $s_{i}$ as a free variable}
Proceeding as in the previous case and choosing $i$ in the range $1 \leq i \leq
q$ and then $s_i$ as the free index, gives 
\begin{align}\label{eq:hypsifree}
  \left[ (-1)^{p+q-1}x \right]^{1-b_i} & \frac{\Gamma(b_i-1)}{\Gamma(1-b_i)}
  \prod_{j} \frac{\Gamma(1-a_j)}{\Gamma(b_i-a_j)}
  \prod_{k \ne i} \frac{\Gamma(b_i-b_k)}{\Gamma(1-b_k)}\\
  & \times \; \pFq{p}{q}
  {\{a_j+1-b_i\}_{1\le j \le p}}
  {2-b_i, \{1-b_k+b_i\}_{1\le k \le q, k\ne i}}
  {x}. \nonumber
\end{align}

\subsection*{Summary}
Assume $p=q+1$ and sum up the series coming from the method of brackets
converging in the common region $|x| > 1$. Rule \ref{rule:freeparams} gives the
analytic continuation
\begin{align}
  {}_{q+1}F_q(x) = \sum_{i=1}^{q+1} (-x)^{-a_i} &
  \prod_{j \ne i} \frac{\Gamma(a_j-a_i)}{\Gamma(a_j)}
  \prod_{k} \frac{\Gamma(b_k)}{\Gamma(b_k-a_i)}\\
  & \times \pFq{q+1}{q}
  {a_i, \{1-b_k+a_i\}_{1\le k \le q}}
  {\{1-a_j+a_i\}_{1\le j \le q+1, j\ne i}}
  {\frac{1}{x}} \nonumber
\end{align}
for the series (\ref{pFqseries}). 

On the other hand, the $q+1$ functions coming from choosing $n$ or $s_i$, $1\le
i\le q$, as the free variables form linearly independent solutions to the
hypergeometric differential equation
\begin{equation}\label{eq:hypdiffeq}
  \prod_{j=1}^{q+1} \left( x \frac{\mathd}{\mathd x} + a_j\right) y
  = \prod_{k=1}^q \left( x \frac{\mathd}{\mathd x} + b_k\right) y
\end{equation}
in a neighborhood of $x=0$. Likewise, the $q+1$ functions (\ref{eq:hyptifree})
coming from choosing $t_i$, $1\le i\le q+1$, as the free variables form linearly
independent solutions to (\ref{eq:hypdiffeq}) in a
neighborhood of $x=\infty$.

\begin{example}
For instance, if $p=2$ and $q=1$ then 
\begin{align}
  \pFq21{a,b}{c}{x} = {} & (-x)^{-a} \frac{\Gamma(b-a)\Gamma(c)}{\Gamma(b)\Gamma(c-a)} \pFq21{a,1-c+a}{1-b+a}{\frac{1}{x}} \\
  + & (-x)^{-b} \frac{\Gamma(a-b)\Gamma(c)}{\Gamma(a)\Gamma(c-b)} \pFq21{b,1-c+b}{1-a+b}{\frac{1}{x}}. \nonumber
\end{align}
This is entry $9.132.1$ of \cite{gr}. On the other hand, the two functions
\begin{equation}
  \pFq21{a,b}{c}{x}, \qquad  x^{1-c} \pFq21{a+1-c,b+1-c}{2-c}{x}
\end{equation}
form a basis of the solutions to the second-order hypergeometric differential equation
\begin{equation}
  \left( x \frac{\mathd}{\mathd x} + a\right)\left( x \frac{\mathd}{\mathd x} + b\right) y
  = \left( x \frac{\mathd}{\mathd x} + c\right) y
\end{equation}
in a neighborhood of $x=0$.
\end{example}

\section{Feynman diagram application}\label{sec-Feynman}

In Quantum Field Theory the permanent contrast between experimental
measurements and theoretical models has been possible due to the development
of novel and powerful analytical and numerical techniques in perturbative
calculations. The fundamental problem that arises in perturbation theory is
the actual calculation of the loop integrals associated to the Feynman
diagrams, whose solution is specially difficult since these integrals
contain in general both ultraviolet (UV) and infrared (IR) divergences.
Using the dimensional regularization scheme, which extends the
dimensionality of space-time by adding a fractional piece $(D=4-2\epsilon )$%
, it is possible to know the behavior of such divergences in terms of
Laurent expansions with respect to the dimensional regulator $\epsilon $
when it tends to zero

As an illustration of the use of method of brackets, the Feynman diagram 
\begin{equation}
  \xygraph{
    !{<0cm,0cm>;<1.5cm,0cm>:<0cm,1.2cm>::}
    !{(0,0)}*+{P_2}="v0"
    !{(1,0)}="v1"
    !{(2,1)}="v2"
    !{(2,-1)}="v3"
    !{(3,1)}*+{P_1}="v4"
    !{(3,-1)}*+{P_3}="v5"
    "v0":"v1" "v2":|*+{a_3}"v3"
    "v1":|*+{a_1}"v2":"v4"
    "v1":|*+{a_2}"v3":"v5"
  }
\end{equation}
considered in \cite{boos1} is resolved.  In this diagram the propagator (or
internal line) associated to the index $a_{1}$ has mass $m$ and the other
parameters are $P_{1}^{2}=P_{3}^{2}=0$ and $P_{2}^{2}=( P_{1}+P_{3})^{2}=s$.
The $D$-dimensional representation in Minkowski space is given by 
\begin{equation}\label{bra4}
  G=\int \frac{d^{D}q}{i\pi ^{D/2}}\frac{1}{\left[ (P_{1}+q)^{2}-m^{2}\right]^{a_{1}}
  \left[ (P_{3}-q)^{2}\right] ^{a_{2}}\left[ q^{2}\right] ^{a_{3}}}.
\end{equation}
In order to evaluate this integral, the Schwinger parametrization of
(\ref{bra4}) is considered (see \cite{itzykson1} for details). This is given
by
\begin{equation}
  G = \frac{(-1)^{-D/2}}{\prod\nolimits_{j=1}^{3}\Gamma(a_{j})} \; H
\end{equation}
with $H$ defined by
\begin{equation}\label{ibfe22}
  H = \int\limits_{0}^{\infty }\int\limits_{0}^{\infty }\int\limits_{0}^{\infty}
  x_{1}^{a_{1}-1}x_{2}^{a_{2}-1}x_{3}^{a_{3}-1}
  \frac{\exp \left(x_{1}m^{2}\right)
  \exp \left(-\frac{x_{1}x_{2}}{x_{1}+x_{2}+x_{3}}s\right)}{\left( x_{1}+x_{2}+x_{3}\right) ^{D/2}}
  \;dx_{1}dx_{2}dx_{3}.
\end{equation}
To apply the method of brackets the exponential terms are expanded as
\begin{equation*}
  \exp \left( x_{1}m^{2}\right) \exp \left(-\frac{x_{1}x_{2}}{x_{1}+x_{2}+x_{3}}s\right)
  = \sum_{n_1,n_2} \phi_{n_1,n_2} \; (-1)^{n_{1}} m^{2n_{1}}s^{n_{2}}
  \frac{x_{1}^{n_{1}+n_{2}}x_{2}^{n_{2}}}{\left( x_{1}+x_{2}+x_{3}\right) ^{n_{2}}},
\end{equation*}
and then (\ref{ibfe22}) is transformed into 
\begin{equation}\label{bra1}
  \sum_{n_1,n_2} \phi_{n_1,n_2}  (-m^{2})^{n_{1}} s^{n_{2}}
  \int\limits_{0}^{\infty }\int\limits_{0}^{\infty }\int\limits_{0}^{\infty}
  \frac{x_{1}^{a_{1}+n_{1}+n_{2}-1}x_{2}^{a_{2}+n_{2}-1}x_{3}^{a_{3}-1}}
  {(x_{1}+x_{2}+x_{3})^{D/2+n_{2}}}\;dx_{1}dx_{2}dx_{3}.
\end{equation}
Further expanding
\begin{equation*}
  \frac{1}{(x_{1}+x_{2}+x_{3})^{D/2+n_{2}}}
  = \sum_{n_3,n_4,n_5} \phi_{n_3,n_4,n_5} \; x_{1}^{n_{3}}x_{2}^{n_{4}}x_{3}^{n_{5}}
  \frac{\bracket{\frac{D}{2}+n_{2}+n_{3}+n_{4}+n_{5}}}{\Gamma(\frac{D}{2}+n_{2})},
\end{equation*}
and replacing into (\ref{bra1}) and substituting the resulting integrals by
the corresponding brackets yields
\begin{align}
  H = {} & \sum_{\{n\}} \phi_{\{n\}} (-1)^{n_{1}} m^{2n_{1}} 
  s^{n_{2}}\frac{\bracket{\tfrac{D}{2}+n_{2}+n_{3}+n_{4}+n_{5}}}{\Gamma(\frac{D}{2}+n_{2})} \\
  & \times \;\bracket{a_{1}+n_{1}+n_{2}+n_{3}}
  \bracket{a_{2}+n_{2}+n_{4}} \bracket{a_{3}+n_{5}}. \nonumber
\end{align}
This bracket series is now evaluated employing Rules \ref{rule:evalbrackets}
and \ref{rule:freeparams}.  Possible choices for free variables are $n_1$,
$n_2$, and $n_4$. The series associated to $n_2$ converges for
$|\tfrac{s}{m^2}| < 1$, whereas the series associated to $n_1, n_4$ converge
for $|\tfrac{m^2}{s}| < 1$. The following two representations
for $G$ follow from here. 

\begin{theorem}
In the region $|\tfrac{s}{m^2}| < 1$,
\begin{equation}\label{eq:feynmanG1}
  H = \eta_2 \cdot \pFq21{a_1+a_2+a_3-\tfrac{D}{2}, a_2}{\tfrac{D}{2}}{\frac{s}{m^2}}
\end{equation}
with $\eta_2$ defined by
\begin{equation*}
  \eta_2 = \left(-m^{2}\right)^{\frac{D}{2}-a_{1}-a_{2}-a_{3}}
  \frac{\Gamma(a_2) \Gamma(a_3) \Gamma\left(a_{1}+a_{2}+a_{3}-\frac{D}{2}\right)
  \Gamma \left( \frac{D}{2}-a_{2}-a_{3}\right) }{\Gamma\left(\frac{D}{2}\right)}.
\end{equation*}
\end{theorem}

\begin{theorem}
In the region $|\tfrac{m^2}{s}| < 1$,
\begin{align}\label{eq:feynmanG2}
  H = {} & \eta_1 \cdot \pFq21{a_1+a_2+a_3-\tfrac{D}{2}, 1+a_1+a_2+a_3-D}{1+a_1+a_3-\tfrac{D}{2}}{\frac{m^2}{s}} \\
  & + \eta_4 \cdot \pFq21{1+a_2-\tfrac{D}{2}, a_2}{1-a_1-a_3+\tfrac{D}{2}}{\frac{m^2}{s}} \nonumber
\end{align}
with $\eta_1$, $\eta_4$ defined by
\begin{align*}
  \eta_1 & = 
  s^{\frac{D}{2}-a_{1}-a_{2}-a_{3}}
  \frac{\Gamma(a_3) \Gamma\left(a_{1}+a_{2}+a_{3}-\frac{D}{2}\right)
  \Gamma\left(\frac{D}{2}-a_{1}-a_{3}\right) \Gamma\left(\frac{D}{2}-a_{2}-a_{3}\right)}
  {\Gamma \left( D-a_{1}-a_{2}-a_{3}\right)}, \\
  \eta_4 & = s^{-a_{2}} \left(-m^{2}\right)^{\frac{D}{2}-a_{1}-a_{3}}
  \frac{\Gamma(a_2) \Gamma(a_3) \Gamma\left(a_{1}+a_{3}-\frac{D}{2}\right)
  \Gamma\left(\frac{D}{2}-a_{2}-a_{3}\right)}{\Gamma\left(\frac{D}{2}-a_{2}\right)}.
\end{align*}
\end{theorem}

These two solutions are now specialized to  $a_1=a_2=a_3=1$.
This situation is specially relevant, since when an arbitrary Feynman
diagram is computed, the indices associated to the propagators are normally
$1$. Then, with
$D=4-2\epsilon$, the equations (\ref{eq:feynmanG1}) and (\ref{eq:feynmanG2}) take the form
\begin{equation}
  H = (-m^2)^{-1-\epsilon} \Gamma(\epsilon-1) \pFq21{1+\epsilon, 1}{2-\epsilon}{\frac{s}{m^2}}
\end{equation}
for $|\tfrac{s}{m^2}| < 1$, as well as
\begin{equation}
  H = s^{-1-\epsilon} \frac{\Gamma(-\epsilon)^2 \Gamma(1+\epsilon)}{\Gamma(1-2\epsilon)} \left(1-\frac{m^2}{s}\right)^{-2\epsilon} 
  - m^{-2\epsilon} \frac{\Gamma(\epsilon)}{\epsilon s} \pFq21{\epsilon, 1}{1-\epsilon}{\frac{m^2}{s}}
\end{equation}
for $|\tfrac{m^2}{s}| < 1$. Observe that these representations both have a pole
at $\epsilon = 0$ of first order (for the second representation, each of the
summands has a pole of second order which cancel each other).

\section{Conclusions and future work}\label{sec-conclusions}

The method of brackets provides a very effective procedure to evaluate definite
integrals over the interval $[0, \infty)$. The method is based on a heuristic
list of rules on the bracket series associated to such integrals. In
particular, a variety of examples that illustrate the power of this method has
been provided. A rigorous validation of these rules as well as a systematic
study of integrals from Feynman diagrams is in progress.

\section*{Acknowledgments}

The first author was partially funded by Fondecyt (Chile), Grant number
$3080029$. The work of the second author was partially funded by NSF-DMS
$0070567$. The last author was funded by this last grant as a graduate student.


\begin{thebibliography}{10}

\bibitem{borwein-2008b}
D.~H. Bailey, J.~M. Borwein, D.~M. Broadhurst, and L.~Glasser.
\newblock Elliptic integral representation of {B}essel moments.
\newblock {\em J. Phys. A: Math. Theor.}, 41:5203--5231, 2008.

\bibitem{boos1}
E.~E. Boos and A.~I. Davydychev.
\newblock A method of evaluating massive {F}eynman integrals.
\newblock {\em Theor. Math. Phys.}, 89:1052--1063, 1991.

\bibitem{borwein-2008a}
J.~M. Borwein and B.~Salvy.
\newblock A proof of a recursion for {B}essel moments.
\newblock {\em Experimental Mathematics}, 17:223--230, 2008.

\bibitem{bailey2006a}
J.~M.~Borwein D.~H.~Bailey and R.~E. Crandall.
\newblock Integrals of the {I}sing class.
\newblock {\em Jour. Phys. A}, 39:12271--12302, 2006.

\bibitem{gonzalez-moll1}
I.~Gonzalez and V.~Moll.
\newblock Definite integrals by the method of brackets. {P}art 1.
\newblock {\em Adv. Appl. Math.}, To appear, 2010.

\bibitem{gonzalez-2007}
I.~Gonzalez and I.~Schmidt.
\newblock Optimized negative dimensional integration method ({NDIM}) and
  multiloop {F}eynman diagram calculation.
\newblock {\em Nuclear Physics B}, 769:124--173, 2007.

\bibitem{gr}
I.~S. Gradshteyn and I.~M. Ryzhik.
\newblock {\em Table of {I}ntegrals, {S}eries, and {P}roducts}.
\newblock Edited by A. Jeffrey and D. Zwillinger. Academic Press, New York, 7th
  edition, 2007.

\bibitem{itzykson1}
C.~Itzykson and J.~B. Zuber.
\newblock {\em Quantum {F}ield {T}heory}.
\newblock World Scientific, Singapore, 2nd edition, 1993.

\bibitem{luis2}
L.~Medina and V.~Moll.
\newblock A class of logarithmic integrals.
\newblock {\em Ramanujan Journal}, 20:91--126, 2009.

\bibitem{palmer-tracy}
J.~Palmer and C.~Tracy.
\newblock Two-dimensional {I}sing correlations: {C}onvergence of the scaling
  limit.
\newblock {\em Adv. Appl. Math.}, 2:329--388, 1981.

\bibitem{skorokhodov1}
S.~L. Skorokhodov.
\newblock Method of analytic continuation of the generalized hypergeometric
  functions $_{p}{F}_{p-1}(a_{1},\cdots,a_{p}; b_{1},\cdots,b_{p-1};z)$.
\newblock {\em Comp. Math. and Math. Physics}, 44:1102--1123, 2004.

\bibitem{skorokhodov2}
S.~L. Skorokhodov.
\newblock Symbolic transformations in the problem of analytic continuation of
  the hypergeometric function $_{p}{F}_{p-1}(z)$ to the neighborhood of the
  point $z=1$ in the logarithmic case.
\newblock {\em Programming and Computer Software}, 30:150--156, 2004.

\end{thebibliography}
\end{document}